\documentclass{emulateapj}

\newcommand{\tco}{$^{13}$CO}

\newcommand{\jo}{$J=1-0$}
\newcommand{\kms}{km s$^{-1}$}
\newcommand{\msun}{$M_{\odot}$}

\shorttitle{Dense Clumps in the IRDC G049.40--00.01}
\shortauthors{Kang et al.}

\begin{document}

\title{Submillimeter Observations of Dense Clumps
       in the Infrared Dark Cloud G049.40--00.01}

\slugcomment{Accepted for publication in ApJ}

\author{\sc Miju Kang$^1$,  
            Minho Choi$^1$,
            John H. Bieging$^2$,
            Jeonghee Rho$^3$,
            Jeong-Eun Lee$^4$,
            and Chao-Wei Tsai$^5$}
\affil{$^1$ Korea Astronomy and Space Science Institute,
            776 Daedeokdaero, Yuseong, Daejeon 305-348, Republic of Korea;
            mjkang@kasi.re.kr.\\
       $^2$ Steward Observatory, University of Arizona,
            933 North Cherry Avenue, Tucson, AZ 85721, USA \\
       $^3$ SOFIA Science Center, USRA/NASA Ames Research Center,
            Moffet Field, CA 94035, USA \\
       $^4$ Department of Astronomy and Space Science, 
            Kyung Hee University, Yongin, Gyeonggi 446-701,
            Republic of Korea \\
       $^5$ Infrared Processing and Analysis Center, 
            California Institute of Technology, Pasadena, CA 91125, USA
            }

\begin{abstract} 

We obtained 350 and 850 \micron\ continuum maps of the infrared dark
cloud G049.40--00.01. Twenty-one dense clumps were  identified within
G049.40--00.01 based on the 350 \micron\ continuum map with an angular
resolution of about 9\farcs6. We present submillimeter continuum maps and
report physical properties of the clumps. The masses of clumps range
from 50 to 600 \msun. About 70\% of the clumps are associated with bright
24 \micron\ emission sources, and they may contain protostars.
The most massive two clumps show extended, enhanced 4.5 \micron\ emission
indicating vigorous star-forming activity.
The clump size-mass distribution suggests
that many of them are forming high mass stars. G049.40--00.01 contains
numerous objects in various evolutionary stages of star formation, from
pre-protostellar clumps to \ion{H}{2} regions.

\end{abstract}



\section{INTRODUCTION}

Massive stars form in cold, dense molecular clouds and in clusters
\citep{Lada03}. Infrared dark clouds (IRDCs) are complexes of cold
($T <$ 20 K), dense ($n >$ 10$^4$ cm$^{-3}$) molecular gas, some of which are
believed to be the progenitors of massive stars and star clusters
\citep{Egan98,Carey98,Rathborne06}. IRDCs were identified as dark
extinction features because the cold dust in IRDCs absorbs the
bright mid-infrared emission of the Galactic plane \citep{Egan98}.
Cold, dense molecular gas and dust in IRDCs were confirmed based on
observations at millimeter and submillimeter wavelengths \citep{Carey98,
Carey00,Rathborne05,Rathborne06}. IRDCs fragment into multiple cores
\citep{Beuther09}. \cite{Battersby10} divided the evolutionary sequence
of IRDCs into four stages from a quiescent clump to an embedded \ion{H}{2}
region by combining millimeter-centimeter continuum data and spectroscopic
data of the HCO$^{+}$ and N$_2$H$^+$ lines. Some IRDCs are thought to be
good targets for investigating the initial conditions of massive star
formation \citep{Beuther07,Rathborne06,Chambers09}.

Recently, \cite{Kang09} catalogued embedded young stellar objects
(YSOs) near W51 using the data from the Galactic Legacy Infrared
Mid-Plane Survey Extraordinaire \cite[GLIMPSE I;][]{Benjamin03} and
the Mid-infrared Imaging Photometer for {\it Spitzer} Galactic plane
Survey \cite[MIPSGAL;][]{Carey09}. A total of 35 YSOs have been found
over the area of $0\fdg15\times0\fdg14$ centered at $l=49\fdg4$ and
$b=0\fdg00$. This region corresponds to MSXDC G049.40--00.01 identified
from the MSX 8 \micron\ data \citep{Simon06a}.  (Hereafter we refer the
region to G049.40--00.01, after dropping the MSXDC label.)  G049.40--00.01
includes three {\it Spitzer} dark clouds catalogued with the GLIMPSE 8
\micron\ data \citep{Peretto09}.  G049.40--00.01 is associated with the
CO emission showing a velocity peak at $V_{\rm LSR}$ = 61 \kms, which
is a part of the ``cluster'' region near the active star forming complex
W51 \citep{Kang10}.
Recent measurements of the distance to W51 gave
$5.41^{+0.31}_{-0.28}$ kpc \citep{Sato10},
$5.1^{+2.9}_{-1.4}$ kpc \citep{Xu09}, and $6.1\pm1.3$ kpc \citep{Imai02}.
Here, we adopt 6 kpc as the distance to G049.40--00.01
(with an uncertainty of $\sim$1 kpc),
for consistency with previous works.
The peak H$_2$ column density estimated from the
\tco\ \jo\ line observations is $1.6 \times 10^{22} \rm ~ cm^{-2}$
\citep{Simon06b}. Two compact \ion{H}{2} regions were identified based
on the {\it Spitzer} data \citep{Phillips08}.

In this paper, we present the results of submillimeter observations of
the IRDC G049.40--00.01 using the Submillimeter High Angular Resolution
Camera II (SHARC-II) at the Caltech Submillimeter Observatory (CSO).
We describe details of the observations and data in Section 2.  Then we
report the results in Section 3 and discuss the physical properties of
the clumps in G49.40--00.01 in Section 4.  We summarize the main results
in Section 5.


\begin{figure*}
\epsscale{0.9}
\plotone{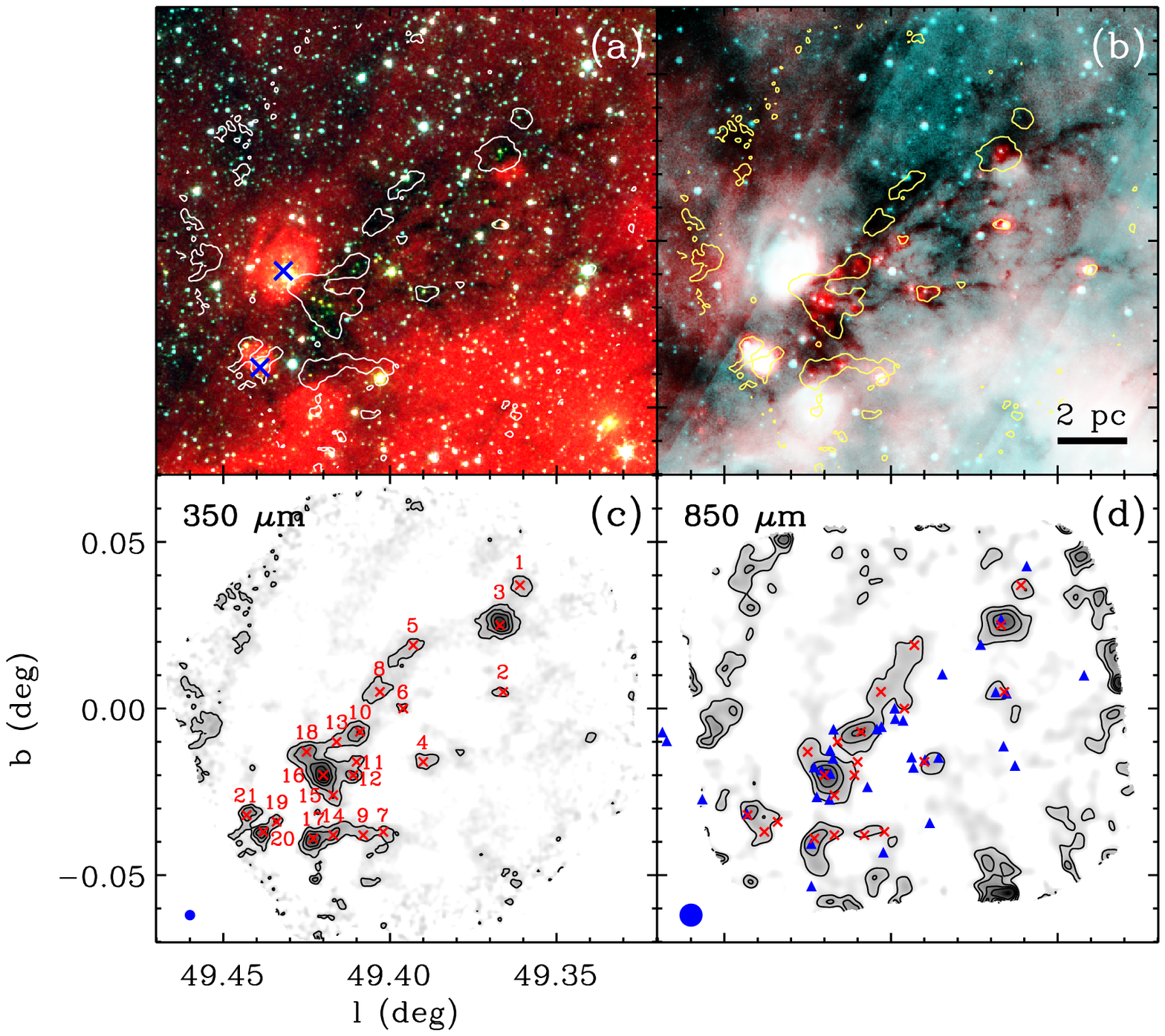}

\caption{Infrared and submillimeter maps of the IRDC G049.40--00.01. (a)
IRAC three-color composite (8.0 \micron\ in red, 4.5 \micron\ in green,
and 3.6 \micron\ in blue) image overlaid with the 4$\sigma$ contour of
the 350 \micron\ map.  Two large crosses mark the compact \ion{H}{2}
regions PR 29 and 30, from north to south \citep{Phillips08}. (b) IRAC
and MIPS two-color composite (8.0 \micron\ in cyan and 24 \micron\ in
red) image. The scale bar indicates 2 pc.  (c) SHARC-II 350 \micron\
image (both gray scale and contours).  The compact sources (clumps)
are labeled and marked by small crosses.  Contour levels are 4, 8,
12, 16, and 20 $\times$ 0.09 (1$\sigma$) Jy beam$^{-1}$.  (d) SHARC-II
850 \micron\ image.  Contour levels are 3, 6, 9, and 12 $\times$ 0.05
(1$\sigma$) Jy beam$^{-1}$.  The crosses mark the positions of the 350
\micron\ clumps. The filled triangles mark the YSOs from \cite{Kang09}.
The FWHM resolutions of the SHARC-II maps are indicated in the bottom
left corners.  The extraneous contours (noisy areas) near the edges of
the SHARC-II maps should be ignored.}

\label{fig1}

\end{figure*}

\section{OBSERVATIONS AND DATA}

The observations were made on 2010 May 10 and 15 at the CSO 10.4 m
telescope near the summit of Mauna Kea, Hawaii. We used the bolometer
camera SHARC-II \citep{Dowell03}. The instrument resolution of SHARC-II is
8\farcs0 at 350 \micron\ and 19\farcs4 at 850 \micron. The Dish Surface
Optimization System was used to correct the dish surface for static
imperfections and deformations due to gravitational forces as the dish
moves in elevation \citep{Leong06}.
We obtained five scans at 350 \micron\
for a total integration time of 50 minutes
in moderate weather ($\tau_{225 \rm GHz} \approx$ 0.046--0.068)
and fourteen scans at 850 \micron\
for a total integration time of 140 minutes
in moderate weather ($\tau_{225 \rm GHz} \approx$ 0.053--0.078).
Pointing and calibration scans
were taken on hourly basis on strong submillimeter sources: Neptune, Arp
220, IRAS 16293--2422, CRL 2688, W75N, and K 3-50.  The data were reduced
using version 2.01-2 of the software package CRUSH \citep[Comprehensive
Reduction Utility for SHARC-II]{Kovacs06}.  The final maps were smoothed
to angular resolutions of FWHM = 9\farcs61 at 350 \micron\ and 23\farcs32
at 850 \micron.

The {\it Spitzer} data presented here are the GLIMPSE and MIPSGAL
data. The GLIMPSE and MIPSGAL are legacy programs covering the
inner Galactic plane at 3.6, 4.5, 5.8, and 8.0 \micron\ with the
Infrared Array Camera \cite[IRAC;][]{Fazio04} and at 24 and 70
\micron\ with the Multiband Imaging Photometer for Spitzer
\cite[MIPS;][]{Rieke04}, respectively. 


\section{RESULTS}

\begin{deluxetable*}{lrr@{ }ccccccl}
\tabletypesize{\scriptsize}
\tablecaption{Submillimeter Continuum Source Parameters}
\tablewidth{0pt}
\tablehead{ & \multicolumn{2}{c}{Peak Position\tablenotemark{a}} & & \multicolumn{2}{c}{Peak Flux\tablenotemark{b}} &
            \colhead{Total Flux} & \colhead{Size\tablenotemark{c}} & \colhead{24 \micron} & \colhead{Associated} \\
            \cline{2-3} \cline{5-6}
            & \colhead{$l$} & \colhead{$b$} & & \colhead{850 $\micron$} & \colhead{350 $\micron$} & \colhead{at 350 $\micron$} &
            & \colhead{Sources\tablenotemark{d}} & \colhead{YSOs\tablenotemark{e}} \\
            \colhead{Source} & \colhead{(deg)} & \colhead{(deg)} & & \colhead{(Jy beam$^{-1}$)} & \colhead{(Jy beam$^{-1}$)} &
            \colhead{(Jy)} & \colhead{($\arcsec \times \arcsec$)} &
            & 
            }
\startdata
  1\dotfill &  49.361 &   0.037 & &   0.30    &0.58  & \ 3.4  $\pm$  0.4 &  21 $\times$  21&   \nodata              &  \\
  2\dotfill &  49.366 &   0.005 & &   0.24    &0.54  & \ 1.6  $\pm$  0.3 &  21 $\times$ \ 8&   2                    & 345, 348 \\
  3\dotfill &  49.367 &   0.025 & &   0.58    &1.75  &  11.8  $\pm$  1.1 &  22 $\times$  19&   2                    & 346 \\
  4\dotfill &  49.390 & --0.016 & &   0.26    &0.54  & \ 2.5  $\pm$  0.2 &  26 $\times$  21&   3                    & 362, 366, 367 \\
  5\dotfill &  49.393 &   0.019 & &   0.23    &0.55  & \ 5.1  $\pm$  1.6 &  53 $\times$  19&   \nodata              &  \\
  6\dotfill &  49.396 & --0.000 & &   0.23    &0.41  & \ 1.2  $\pm$  0.1 &  16 $\times$  15&   1                    &  \\
  7\dotfill &  49.401 & --0.036 & &   \nodata &0.54  & \ 1.6  $\pm$  0.2 &  22 $\times$  16&   \nodata              &  \\
  8\dotfill &  49.403 &   0.005 & &   0.27    &0.56  & \ 4.4  $\pm$  0.6 &  48 $\times$  23&   \nodata              &  \\
  9\dotfill &  49.408 & --0.038 & &   0.19    &0.68  & \ 2.3  $\pm$  0.3 &  29 $\times$  13&   1                    &  \\
 10\dotfill &  49.409 & --0.007 & &   0.48    &1.09  & \ 5.5  $\pm$  1.1 &  21 $\times$  18&   3                    &  \\
 11\dotfill &  49.410 & --0.016 & &   \nodata &0.63  & \ 1.4  $\pm$  0.2 &  16 $\times$  11&   \nodata              &  \\
 12\dotfill &  49.412 & --0.020 & &   \nodata &0.99  & \ 2.2  $\pm$  0.1 &  20 $\times$  11&   1                    &  \\
 13\dotfill &  49.416 & --0.010 & &   \nodata &0.57  & \ 1.4  $\pm$  0.3 &  21 $\times$  11&   1                    &  \\
 14\dotfill &  49.418 & --0.038 & &   \nodata &0.88  & \ 3.4  $\pm$  0.3 &  29 $\times$  13&   1                    &  \\
 15\dotfill &  49.417 & --0.027 & &   \nodata &0.70  & \ 2.1  $\pm$  0.5 &  14 $\times$  13&   \nodata              &  \\
 16\dotfill &  49.420 & --0.021 & &   0.65    &1.96  &  13.3  $\pm$  0.8 &  27 $\times$  20&   3                    & 409, 414, 419 \\
 17\dotfill &  49.423 & --0.040 & &   0.41    &1.44  & \ 5.6  $\pm$  0.8 &  24 $\times$  17&   1                    & 421 \\
 18\dotfill &  49.425 & --0.013 & &   0.27    &1.02  & \ 4.3  $\pm$  0.3 &  33 $\times$  16&   \ 1 \tablenotemark{f}&  \\
 19\dotfill &  49.434 & --0.034 & &   0.21    &0.72  & \ 1.2  $\pm$  0.3 &  13 $\times$ \ 7&   \nodata              &  \\
 20\dotfill &  49.439 & --0.038 & &   0.26    &1.31  & \ 3.4  $\pm$  0.9 &  14 $\times$  12&   \ 1 \tablenotemark{f}&  \\
 21\dotfill &  49.443 & --0.032 & &   0.40    &0.94  & \ 2.3  $\pm$  0.4 &  21 $\times$  17&   1                    & 438
\enddata
\tablenotetext{a}{Peak position in the 350 \micron\ image.}
\tablenotetext{b}{Uncertainties are 0.05 and 0.09 Jy beam$^{-1}$
                  at 850 and 350 \micron, respectively.}
\tablenotetext{c}{Deconvolved FWHM size of each clump in the 350 \micron\ map.
                  The first number is the size along the major axis
                  (longest diameter),
                  and the second number is the size
                  along the direction perpendicular to the major axis.}
\tablenotetext{d}{Number of 24 \micron\ point sources
                  within the size of each clump.}
\tablenotetext{e}{YSO number in Table 2 of \cite{Kang09}.}
\tablenotetext{f}{Associated with an H {\scriptsize II} region.}

\label{table1}

\end{deluxetable*}

Figure \ref{fig1} shows the maps of the {\it Spitzer} IRAC, MIPS, and
SHARC-II 350 and 850 \micron\ toward the IRDC G049.40--00.01. In the
IRAC composite map (Figure \ref{fig1}a), a dark filamentary structure
is seen in absorption against the background emission.  The bright
features in the IRAC composite map are two \ion{H}{2} regions and bright
diffuse emission from the active star-forming region W51 to the south.
The spatial distribution of the dark filamentary structure agrees well
with the distribution of the 350 \micron\ emission.  Figure 1b shows the
two-color composite (IRAC 8.0 \micron\ in cyan and MIPS 24 \micron\ in
red) image, overlaid with the 350 \micron\ emission map.  Some regions of
G049.04--00.01 are still dark at 24 \micron, implying low temperature and
high column density.  There are also many bright 24 \micron\ point-like
sources located within the IRDC.  The dark features of the IRDC seen in
the {\it Spitzer} maps appear in emission in the 350 and 850 \micron\
maps (Figures 1c and 1d) because their cold thermal dust emission peaks
at millimeter/submillimeter wavelengths.  The bright clumps in the 350
\micron\ map are well aligned with the 24 \micron\ peaks, which suggests
that the 24 \micron\ sources are central stars of these dense clumps.
The distribution of 850 \micron\ emission is very similar to that of
350 \micron\ emission except the details unresolved with the relatively
large beam of the 850 \micron\ map.  The maximum flux densities are 1.96
$\pm$ 0.09 and 0.64 $\pm$ 0.05 Jy beam$^{-1}$ at 350 and 850 \micron,
respectively.

A total of 21 sources were identified as compact clumps for which peak
flux densities are greater than the 4$\sigma$ level in the 350 \micron\
continuum map by eye.  When a peak is located close to another one, it
is considered an independent peak if the peak intensity is higher than
the intensity at the interface between them by 1$\sigma$ level or larger
and if their separation is larger than a beam size.  Table \ref{table1}
lists the peak position, peak flux densities, 350 \micron\ total flux
density, size, number of associated 24 \micron\ sources, and associated
YSOs of each clump.  The total flux density at 350 \micron\ is measured
from the emission in the circumscribed box of 2$\sigma$ contour.
In crowded areas, the saddle point between adjacent clumps is used to
limit the area for total flux measurements.
The size represents the FWHM, deconvolved with the beam. 
In crowded areas, the intensity at the saddle point between adjacent clumps
can be higher than the half-maximum intensity. In this case, we give the
distance to the saddle point from the peak position as a size (in the
direction to the adjacent clump). Column 8 lists
the number of all 24 \micron\ point sources detected with significant
signal-to-noise ratios (S/N $>$ 4).  Column 9
lists the YSOs in \cite{Kang09}. They used a detection criterion of S/N
$>$ 7 at 24 \micron\ to identify YSOs.  The peak fluxes at 850 \micron\
are measured only for the clumps having 850 \micron\ peaks within the
size boundaries of the 350 \micron\ sources as described above.
Table \ref{table2} lists the 850 $\micron$ total flux densities.
For many clumps the 850 \micron\ total flux
cannot be measured individually because of crowding.

\begin{deluxetable}{lr@{ }r@{ }r}
\tabletypesize{\scriptsize}
\tablecaption{Total flux at 850 $\micron$}
\tablewidth{0pt}
\tablehead{
&\multicolumn{3}{c}{Total Flux}\\
\colhead{Sources} & \multicolumn{3}{c}{(Jy)}
}
\startdata
1\dotfill                      & 0.30 & $\pm$ & 0.05 \\
2\dotfill                      & 0.29 & $\pm$ & 0.13 \\
3\dotfill                      & 1.03 & $\pm$ & 0.15 \\
4\dotfill                      & 0.26 & $\pm$ & 0.05 \\
5, 6, 8\dotfill                & 0.98 & $\pm$ & 0.18 \\
7, 9\dotfill                   & 0.22 & $\pm$ & 0.12 \\
10, 13\dotfill                 & 0.79 & $\pm$ & 0.15 \\
11, 12, 15, 16, 18\ldots\ldots & 1.44 & $\pm$ & 0.18 \\
14, 17\dotfill                 & 0.73 & $\pm$ & 0.15 \\
19, 20, 21\dotfill             & 0.51 & $\pm$ & 0.14
\enddata
\tablecomments{For crowded areas,
               the total flux densities of each area are listed.}
\label{table2}
\end{deluxetable}


\section{DISCUSSION}

\subsection{Physical Parameters}

Continuum emission at 350 \micron\ is sensitive to dust temperature,
and the mass of molecular gas estimated from the 350 \micron\ flux density
depends on the assumed dust temperature
that varies with the presence or absence of central heating
by embedded protostars.
The clumps without detectable
24 \micron\ sources may be pre-protostellar (harboring no protostar).
\cite{Hennemann09} derived dust temperatures of 22 and 15 K for cores with
and without 24 \micron\ source, respectively. \cite{Stutz10} found that
the dust temperatures are $\sim$17.7 K near the protostar and $\sim$10.6
K for the starless core in the Bok globule CB 244. Pre-protostellar and
protostellar cores in the IRDC G011.1--0.12 have core temperatures of 22 K
\citep{Henning10}. \cite{Wilkock11} derived temperatures of 8--11 K at the
center of the cores and 18--28 K at the surface using radiative transfer
models of IRDC seen in {\it Herschel} observations.  \cite{Peretto10}
reported that IRDCs are not isothermal, showing that the dust temperature
decreases significantly within IRDCs, from background temperatures of
20--30 K to minimum temperatures of 8--15 K within the clouds.  In this
paper, we assume a dust temperature of 15 K for simplicity and for ease of
comparison with \cite{Kauffmann10} (see Section 4.3).

The masses of the submillimeter clumps in G049.40--00.01
were calculated using the 350 \micron\ flux densities.
The mass can be estimated by
  $ M = {F_\nu D^2}/{\kappa_\nu B_\nu(T_d)}$, 
where $F_\nu$ is the flux density at 350 \micron, $D$ is the distance
to the source, $B_\nu$ is the Planck function, $T_d$ is the dust
temperature, and $\kappa_\nu$ is the dust mass opacity.  The value of
$\kappa_\nu$ at 350 \micron\ used to calculate the masses is $5.91$
cm$^2$ g$^{-1}$, from the coagulated dust model with thin ice mantles
of \cite{Ossenkopf94}. Table \ref{table3} lists the total mass within
the 2$\sigma$ contour. The representative masses listed in Table \ref{table3}
were calculated using a uniform temperature of 15 K. These masses can be
underestimates for colder clumps (those without a 24 \micron\ source)
and overestimates for hotter clumps (those with 24 \micron\ sources).
Different assumptions on the dust temperature can increase or decrease
the mass estimate by a factor of $\sim$3 (see the footnote of Table
\ref{table3}).  The uncertainty of the mass caused by the uncertainty of
the distance is $\sim$30\%.  The sum of the clump masses is $\sim$3,600
\msun.

\begin{deluxetable}{lr@{ }r@{ }rr@{ }r@{}rc}
\tabletypesize{\scriptsize}
\tablecaption{Properties of the Clumps}
\tablewidth{0pt}
\tablehead{
&\multicolumn{3}{c}{M$_{\rm total}$\tablenotemark{a}}
&\multicolumn{3}{c}{M$_{\rm FWHM}$\tablenotemark{b}}
&\colhead{Concentration}\\
\colhead{Source}
&\multicolumn{3}{c}{($M_\odot$)}
&\multicolumn{3}{c}{($M_\odot$)}
&
}
\startdata
  1\dotfill &    154& $\pm$ & 18 &  149& $\pm$  & 27 &  0.65 $\pm$ 0.18\\
  2\dotfill &     73& $\pm$ & 14 &   54& $\pm$  & 20 &  0.73 $\pm$ 0.44\\
  3\dotfill &    536& $\pm$ & 50 &  354& $\pm$  & 75 &  0.77 $\pm$ 0.05\\
  4\dotfill &    113& $\pm$ &  9 &  127& $\pm$  & 14 &  0.65 $\pm$ 0.21\\
  5\dotfill &    232& $\pm$ & 73 &  226& $\pm$  &109 &  0.78 $\pm$ 0.54\\
  6\dotfill &     54& $\pm$ &  5 &   58& $\pm$  &  7 &  0.57 $\pm$ 0.17\\
  7\dotfill &     73& $\pm$ &  9 &   88& $\pm$  & 14 &  0.50 $\pm$ 0.08\\
  8\dotfill &    200& $\pm$ & 27 &  217& $\pm$  & 41 &  0.55 $\pm$ 0.09\\
  9\dotfill &    104& $\pm$ & 14 &   99& $\pm$  & 20 &  0.59 $\pm$ 0.08\\
 10\dotfill &    250& $\pm$ & 50 &  182& $\pm$  & 75 &  0.72 $\pm$ 0.13\\
 11\dotfill &     64& $\pm$ &  9 &   60& $\pm$  & 14 &  0.50 $\pm$ 0.07\\
 12\dotfill &    100& $\pm$ &  5 &  103& $\pm$  &  7 &  0.64 $\pm$ 0.17\\
 13\dotfill &     64& $\pm$ & 14 &   64& $\pm$  & 20 &  0.39 $\pm$ 0.06\\
 14\dotfill &    154& $\pm$ & 14 &  171& $\pm$  & 20 &  0.69 $\pm$ 0.23\\
 15\dotfill &     95& $\pm$ & 23 &   90& $\pm$  & 34 &  0.62 $\pm$ 0.13\\
 16\dotfill &    604& $\pm$ & 36 &  455& $\pm$  & 54 &  0.67 $\pm$ 0.03\\
 17\dotfill &    254& $\pm$ & 36 &  242& $\pm$  & 54 &  0.71 $\pm$ 0.14\\
 18\dotfill &    195& $\pm$ & 14 &  231& $\pm$  & 20 &  0.54 $\pm$ 0.05\\
 19\dotfill &     54& $\pm$ & 14 &   61& $\pm$  & 20 &  0.62 $\pm$ 0.08\\
 20\dotfill &    154& $\pm$ & 41 &  139& $\pm$  & 61 &  0.64 $\pm$ 0.04\\
 21\dotfill &    104& $\pm$ & 18 &  131& $\pm$  & 27 &  0.65 $\pm$ 0.08
\enddata
\tablecomments{The mass is based on the total flux density at 350 \micron\
               assuming $T_d$ = 15 K.
               If $T_d$ = 10 K, multiply by 4.1.
               If $T_d$ = 20 K, multiply by 0.47.}
\tablenotetext{a}{Mass within the 2$\sigma$ contour.}
\tablenotetext{b}{Mass within the FWHM boundary with the opacity scaled by 1/1.5.}

\label{table3}

\end{deluxetable}

\subsection{Comparison with Simple Models}

\begin{figure}
\epsscale{1.0}

\plotone{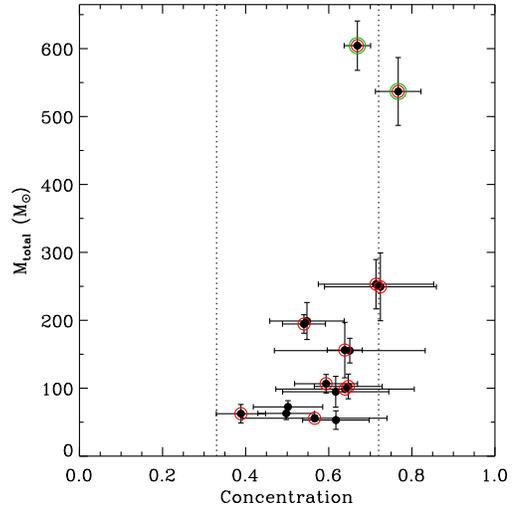}

\caption{Mass-concentration diagram for the clumps detected in
the 350 \micron\ map.
Clumps with large ($>$ 0.2) uncertainties in concentration are omitted.
Red and green circles represent
clumps associated with 24 \micron\ sources and green fuzzies, respectively
(see Section 4.3).
The vertical dotted lines ($C$ = 0.33 and 0.72) represent
the limiting cases of stable Bonnor-Ebert spheres \citep{Johnstone00}.}

\label{fig2}

\end{figure}

The clumps in G049.40--00.01 could be produced by gravitational
fragmentation, and some simple quantities can be calculated to check
the consistency.  The average surface density is $\sim$0.04 g cm$^{-2}$
for the area within the 1$\sigma$ contour in the 350 \micron\ map.
With this surface density and the assumed dust temperature of 15 K,
the critical wavelength of an isothermal, infinite, self-gravitating
cylinder is $\lambda_c \approx 0.3$ pc, and the corresponding
critical mass is $M_c \approx 5$ \msun\ \citep{Hartmann02,Larson85}.
For comparison, the projected distance between the nearest neighbours
of clumps in G049.40--00.01 lies between 0.4 and 2 pc.  The mean
clump separation is 0.9 pc, and the average clump mass is 170 \msun.
Many filamentary IRDCs fragment into clumps with a similar fragmentation
scale \citep{Henning10,Miettinen10}.  The difference between the
calculated critical mass and the measured average mass probably
indicates that the fragmentation history of G049.40--00.01 is more
complicated than the fragmentation of an idealized simple cylinder.
Previously suggested possibilities include an extra support by turbulence,
changes of cloud temperature, and a compression by external forces
\citep{Onishi98,Miettinen10}.

To study further physical conditions of the clumps, we compare the clumps
to a static cloud model.  The Bonnor-Ebert model describes the simplest
self-gravitating pressure-confined isothermal sphere in a hydrostatic
equilibrium \citep{Ebert55, Bonnor56}.  The degree of self-gravitation
within each clump can be estimated by calculating the degree of
concentration of each clump.  The concentration can be defined by $C =
1 - R_\Sigma$, where $R_\Sigma$ is the ratio of average to central column
density (see equation (4) of \cite{Johnstone00}).  Small concentrations
($C < 0.33$) imply uniform-density non-self-gravitating objects, and
large values ($C > 0.72$) imply critically self-gravitating objects
\citep{Johnstone00}.
The concentrations of the submillimeter clumps
are listed in Table \ref{table3}.
Figure \ref{fig2} shows the concentration of the clumps against the
clump mass.  Most clumps in G049.40--00.01 have concentrations between
0.33 and 0.72.  Clump 3 ($C = 0.77 \pm 0.05$) is more concentrated than
what is permitted by stable Bonnor-Ebert spheres.  In general, massive
clumps seem to have high concentrations.  It is interesting to note
that the four highest-concentration clumps are also the highest-mass
ones, and they are all associated with 24 \micron\ sources, indicating
ongoing star formation.  Two of them also show signs of shocked gas (see
Section 4.3).  For the rest of the clumps, there is no clear difference
between those with and without 24 \micron\ sources.

\subsection{High-mass Star Formation}

Massive stars are expected to be formed in molecular clouds with a large
mass concentrated in a relatively small volume.
\cite{Kauffmann10} suggested a threshold for massive star formation
by comparing clouds with and without massive star formation:
$m_{\rm KP} = 870\ M_\odot\ r^{1.33}$,
where $r$ is the effective radius
(half of the geometric mean of the FWHM sizes in Table 1) in pc.
Clouds/clumps more massive than
this threshold seem to form massive stars. \cite{Parmentier11} provided
an explanation for the Kauffmann-Pillai threshold by calculating the
probable mass of the most massive star formed in model clumps having
power-law density profiles.

\begin{figure}
\epsscale{1.0}

\plotone{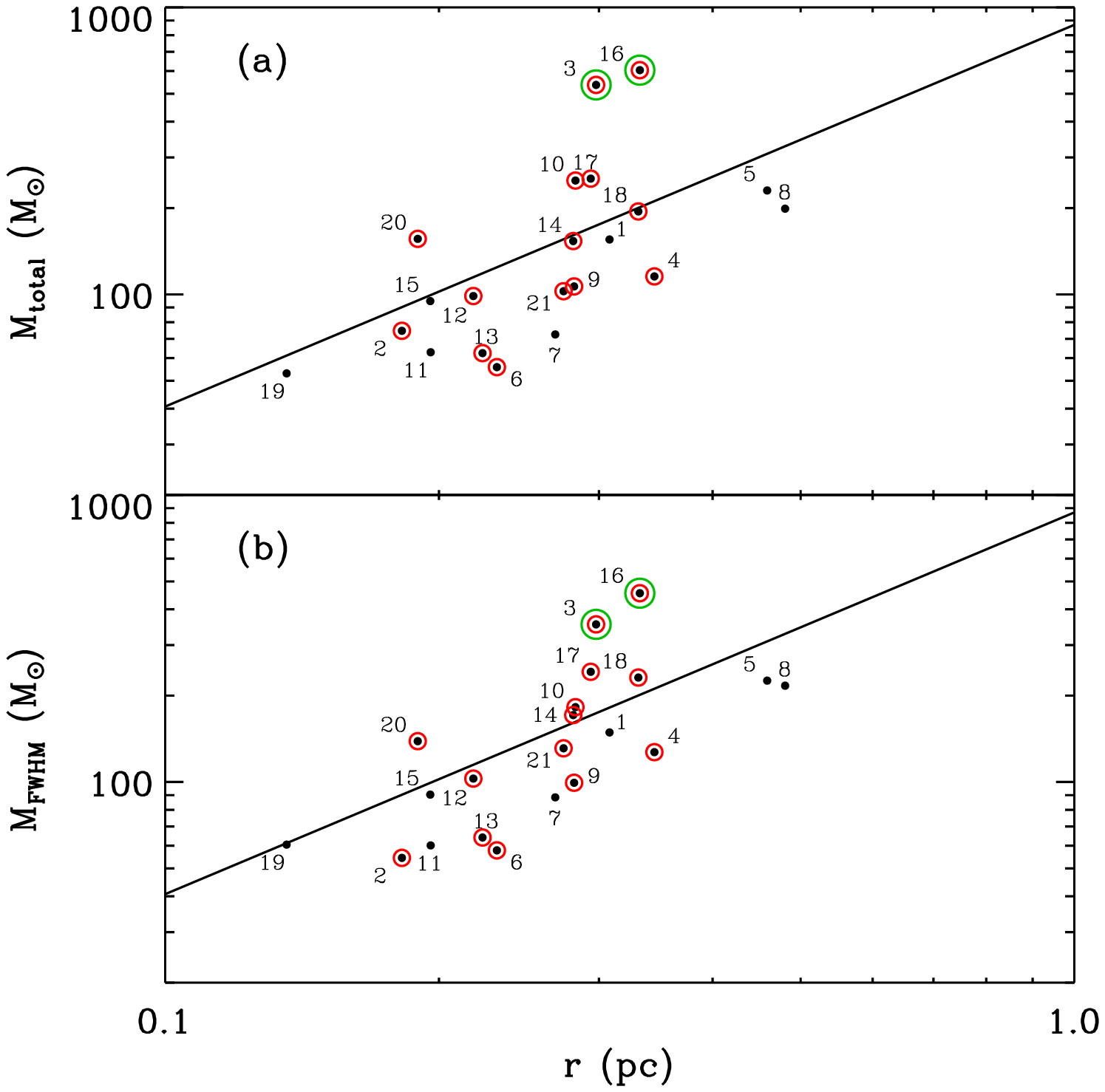}

\caption{
Mass-size diagrams for the clumps detected in the 350 \micron\ map.
(a) Mass-size diagram with the total mass given in Table \ref{table3}.
(b) Mass-size diagram with the mass
estimated from the flux density within the FWHM boundary
(and with the dust opacity scaling factor applied,
as suggested by \cite{Kauffmann10}).
Clumps associated with 24 \micron\ sources are marked by red circles.
Two large green circles mark the ``green fuzzies''.
The solid line represents the mass-size threshold for massive star formation,
$m_{\rm KP}$, proposed by \cite{Kauffmann10}.}

\label{fig3}

\end{figure}

Figure \ref{fig3} shows the mass-size relation for
the clumps in G049.40--00.01. 
\cite{Kauffmann10} assumed a
dust temperature of 15 K.
They used a dust opacity scaled by a factor of 1/1.5
(to match the masses estimated from dust emission and extinction)
and a mass correction factor of 1/$\ln2$
(to convert the total mass to the mass contained in the FWHM),
and these factors cancel out almost exactly.
Figure \ref{fig3}a shows the relation
between the total mass and the effective radius.
Figure \ref{fig3}b shows the relation
between the mass contained within the FWHM boundary
(with the opacity scaled by 1/1.5
to make a proper comparison with \cite{Kauffmann10})
and the effective radius. The masses are listed in Table \ref{table3}.
All the clumps are distributed near $m_{\rm KP}$,
within a factor of $\sim$3.5.

Association with bright 24 \micron\ point-like sources
is an indicator of star formation
because 24 \micron\ emission traces warm dust
heated by the material accreting onto a central protostar.
Fourteen clumps out of 21
have 24 \micron\ point-like sources within their extent.
The most massive ones (clumps 3 and 16) even show extended, enhanced 4.5
\micron\ emission called ``green fuzzies'' or extended green objects,
which indicate shocked gas \citep{Chambers09, Cyganowski08}.  Extended
4.5 \micron\ emission also indicates vigorous star-forming activity.
In the G049.40--00.01 region, two compact \ion{H}{2} regions (PR 29
and 30 in Figure \ref{fig1}a) were identified based on {\it Spitzer}
data \citep{Phillips08}.  PR 29 is more extended than PR 30 in the
infrared maps (Figures 1a and 1b) and is not directly associated with a
submillimeter clump.  PR 29 seems to be a blister type \ion{H}{2} region
formed on the cloud surface near clump 18.  In contrast, PR 30 is closely
associated with the submillimeter clump 20, which suggests that it is
still embedded in the dense cloud.  PR 30 may be less evolved than PR 29.
Clumps 3, 16, and 20 (clumps associated with ``green fuzzies'' or compact
\ion{H}{2} region) are located well above $m_{\rm KP}$, and this fact
seems to corroborate the threshold for massive star formation suggested
by \cite{Kauffmann10}.

Several clumps are dark at 24 \micron,
and they may be clumps either in the pre-protostellar phase
or containing protostars of undetectably low luminosity.
These clumps reside below $m_{\rm KP}$ in Figure \ref{fig3}
and may not be dense enough to form stars (yet).
Therefore, objects in various stages of star formation
(from pre-protostellar cores, protostars, to \ion{H}{2} regions)
are distributed in and around the IRDC G049.40--00.01.  

Spectral observations would be required to definitively determine
the evolutionary status of the clumps. Recent multi-wavelength, high angular
resolution studies on IRDCs focused on the chemical, kinematic, and
physical properties of the initial conditions for massive star formation.
\cite{Pillai2011} investigated secondary massive cold cores in the
vicinity of \ion{H}{2} regions based on the line (NH$_2$D, NH$_3$, and
HCO$^+$) and millimeter continuum observations.  They showed that the
cores in the earliest stage of massive star formation are cold, dense, and
highly deuterated ([NH$_2$D/NH$_3$] $>$ 6\%).  \cite{Devine11} found an
anti-correlation between NH$_3$ and CCS toward the IRDC G19.30+0.07 based
on interferometric observations at 22 GHz.  The different evolutionary
states of the young objects in G049.40--00.01 can be investigated in
detail by obtaining high-resolution interferometric data in the future.


\section{SUMMARY}

We observed the IRDC G049.04--00.01 in the 350 and 850 \micron\ continuum
with the SHARC-II bolometer camera.
The dark features in the infrared images
have a good agreement with the emission structure in the submillimeter images.
Twenty-one clumps were identified based on the 350 \micron\
continuum map, and the mass of each clump ranges from 50 to 600 \msun.
The majority of these clumps are associated with bright 24 \micron\
emission sources, indicating star-forming activity.  The most massive
clumps (clumps 3 and 16) show extended, enhanced emission in the IRAC
4.5 \micron\ image.  All the clumps are distributed near the threshold
for massive star formation. The IRDC G049.04--00.01 contains objects in
various evolutionary stages of star formation.

\acknowledgments 

We thank Hiroko Shinnaga and Michael Dunham for helpful discussions.
M. K. and M. C. were supported by the Core Research Program of the
National Research Foundation of Korea (NRF) funded by the Ministry
of Education, Science and Technology (MEST) of the Korean government
(grant number 2011-0015816).  J.-E. L. was supported by the Basic Science
Research Program through NRF funded by MEST (grant number 2011-0004781).
This work is based in part on observations made with the {\it Spitzer
Space Telescope}, which is operated by the Jet Propulsion Laboratory,
California Institute of Technology, under a contract with NASA.
Caltech Submillimeter Observatory (CSO) is supported through NSF grant
AST-0540882.


\end{document}